# A model colloidal gel for coordinated measurements of force, structure, and rheology

Lilian C. Hsiao[1,†], Kathryn A. Whitaker[2,†], Michael J. Solomon[1] and Eric M. Furst[2,*]

[1]University of Michigan, Department of Chemical Engineering, Ann Arbor, Michigan 48197, USA

[2]University of Delaware, Department of Chemical Engineering, Center for Molecular and Engineering Thermodynamics, Newark, Delaware 19716, USA

We introduce a model gel system in which colloidal forces, structure, and rheology are measured by balancing the requirements of rheological and microscopy techniques with those of optical tweezers. Sterically stabilized poly(methyl methacrylate) (PMMA) colloids are suspended in cyclohexane (CH) and cyclohexyl bromide (CHB) with dilute polystyrene serving as a depletion agent. A solvent comprising of 37% weight fraction CH ($w_{CH}$ = 0.37) provides sufficient refractive index contrast to enable optical trapping, while maintaining good confocal imaging quality and minimal sedimentation effects on the bulk rheology. At this condition, and at a depletant concentration $c$ = 8.64 mg/mL ($c/c^*$ = 0.81), results from optical trapping show that 50% of bonds rupture at (3.3 ± 0.5) pN. The linear strain-dependent elastic modulus of the corresponding gel ($\phi$ = 0.20) is G' = (1.8 ± 0.6) Pa, and the mean contact number of the particles in the gel structure is $\langle z \rangle$ = 5.4 ± 0.1. These structural and rheological parameters are similar to colloidal gels that are weakly aggregating and cluster-like. Thus, the model gel yields a concomitant characterization of the interparticle forces, microstructure, and bulk rheology in a single experimental system, thereby introducing the simultaneous comparison of these experimental measures to models and simulations.

## I. INTRODUCTION

The solid-like elastic properties in colloidal gels, which have significant technological and scientific applications [Lu et al. (1997); Drury and Mooney (2003); Huh et al. (2007); Helgeson et al. (2012)], are a result of the three-dimensional stress-bearing structures that are formed by physical or chemical attraction between the individual components of the network [Vermant and Solomon (2005); Zaccarelli (2007)]. The short range of the interactions used to induce gelation and distinct separation of bound and unbound states between particles is analogous to physical or chemical bonds between molecules and macromolecules. Bond rupture occurs when a sufficiently large force is applied to overcome the deep potential wells. This phenomenon is known as yielding, in which structural evolution such as local rearrangement [Park and Ahn (2013)], bond and cage breakage [Koumakis and Petekidis (2011)], and flow anisotropy occur [Rajaram and Mohraz (2010); Hoekstra et al. (2003)]. These complex structural changes are manifest in the yield stress and other rheological measures such as the nonlinear creep compliance [Gopalakrishnan and Zukoski (2007)] and the relaxation modulus [Yin and Solomon (2008)]. The microscopic origins of rheological phenomenon are commonly probed through a combination of rheometry and characterization of the local structure and dynamics, either by direct visualization [Schall et al. (2007); Masschaele et al. (2011); Cheng et al. (2011); Emady et al. (2013); Lee and Furst (2008)] or by scattering methods [Vermant et al. (1998); Varadan and Solomon (2001); Maranzano and Wagner (2001); Mohraz and Solomon (2005); Ramakrishnan et al. (2005); Reddy et al. (2012)].

Micromechanical models seek to predict the rheological properties of gels in shear flow based on the transient microstructure and were originally developed for fractal gels at low volume fractions [Shih et al. (1990); West et al. (1994); Potanin (1994)]. Current models show system-specific discrepancies in the power-law scaling between nonlinear rheology and interparticle attraction [Studart et al. (2011); Hess and Aksel (2013)]. Despite the importance of the particle-level forces that give rise to gelation and the viscoelastic response in shear flow, none of these scaling laws accounts for the spectrum of forces and relaxation time scales responsible for bond rupture [Evans (1998); Swan et al. (2012)]. Instead, models rely on the athermal force limit of rupture as an input [Studart et al. (2011)], where the maximum slope of the potential is used to predict the force required for particles to escape from the potential minimum. For instance, the athermal limits of the van der Waals attraction and depletion interaction encountered in many colloidal systems can be analytically obtained from classical derivations at dilute limits [Asakura and Oosawa (1958); Russel et al. (1989)]. Despite its simplicity, the assumption of a singular force that is responsible for bond rupture is problematic because of the stochastic nature of bond breaking [Kramers (1940)]. External fields (such as gravity or an applied strain acting on a colloidal gel) can lower the energy barrier [Fielding et al. (2000); Manley et al. (2004)], making it possible for bonds to rupture at a force below the athermal limit. A realistic micromechanical model should account for the effect of field-induced thermal fluctuations on the probability of bond rupture.

Direct manipulation of colloids using optical and magnetic tweezers for instance, is a suitable method to determine the distribution of particle-level forces responsible for linear and nonlinear rheology [Crocker et al. (1998); Sriram et al. (2009); Rich et al. (2011); Swan et al. (2012); Pantina and Furst (2008); Furst and Pantina (2007); Pantina and Furst (2006); Pantina and Furst (2005)]. Nevertheless, a model depletion-driven system that allows for direct force measurements has not been reported in the literature due to the challenges in satisfying a number of experimental limitations, particularly if force measurements are sought in a system in which rheology and microstructure can also be measured. Characterization of 3D gel microstructure using confocal microscopy requires a close refractive index match between the colloids and the solvent. In addition, systems are commonly density-matched to minimize sedimentation effects on microstructure and rheology. Most studies incorporate a specific solvent mixture to meet these requirements [Royall et al. (2003); Dibble et al. (2006); Chan and Mohraz (2012)]. The low refractive index contrast between the particles and the solvent in this type of model system prevents the measurement of thermal rupture forces. Here, we present a colloidal system in which the interparticle forces generated through a short-ranged depletion interaction are directly measured with laser tweezers, concomitant with characterizations of the microstructure and viscoelastic modulus of the gel at a specific solvent composition. The three challenges of imaging quality, sedimentation, and optical trapping strength are balanced by a systematic investigation of a variety of solvent mixtures spanning a large range of refractive index contrast ratios and density differences.

## II. EXPERIMENTAL METHODS

A. Colloidal synthesis and gelation at various solvent compositions

Monodisperse poly(methyl methacrylate) (PMMA) colloids (diameter $2a = 1.15$ μm ± 3% and 2.7 μm ± 3%; density $\rho_p = 1.19$ g/mL; refractive index $n_p = 1.488$) stabilized by a grafted layer of poly(12-hydroxystearic acid) (PHSA) [Antl et al. (1986)] are used in this study. The thickness of the PHSA steric layer is 10 – 14 nm [Campbell and Bartlett (2002)]. PMMA colloids are dyed with fluorescent Nile Red to allow for direct visualization with confocal microscopy. Gelation is induced in particles of $2a = 1.15$ μm by addition of monodisperse polystyrene (molecular weight $M_w = 900,000$ g/mol, $M_w/M_n = 1.10$) at a fixed concentration, $c = 8.64$ mg/mL. For microstructural and rheological characterization, gels of intermediate volume fraction (ϕ = 0.20) are used. Laser tweezer experiments are performed on a dilute system containing the larger PMMA particles ($2a = 2.7$ μm). The different particle sizes used for different characterization techniques are reconciled by dimensional scalings appropriate for Brownian systems [Mewis and Wagner (2012)].

Mixtures of cyclohexane (CH) and cyclohexyl bromide (CHB) at various solvent compositions are chosen for this study. Combining CH with CHB provides a sufficiently large refractive index contrast so that the traps generated by the laser tweezers are strong enough to measure rupture forces. (The most common solvents used to achieve buoyancy-matching in studies of colloidal suspensions, CHB and decalin, do not provide sufficient refractive index contrast for optical trapping.) These mixtures result in a density difference between the particles and the solvent, but we show that there are conditions where the sedimentation does not significantly affect the microstructural characterization or rheology. The solvents are washed with deionized water and filtered through a 0.2 μm polytetrafluoroethylene syringe filter prior to use. Table I shows the full range of solvent mixtures used in this study along with their physical properties, where $w_{CH}$ and $w_{CHB}$ are the mass fractions of CH and CHB in the solvents, respectively. The Péclet number, defined as $Pe = 4\pi\Delta\rho g a^4 / 3kT$, characterizes the ratio of single-particle sedimentation to that of Brownian diffusion in the rheological and microstructural characterization experiments [Russel et al. (1989)]. Here $\Delta\rho = \rho_s - \rho_p$, $\rho_s$ is the measured density of the mixed solvent (Anton Paar DMA 4500 M), $\rho_P$ is the density of the particle, $k$ is the Boltzmann constant, and $2a = 1.15$ μm, the size of the colloids used for microstructural and rheological characterization. A large density mismatch also corresponds to a large refractive index contrast, $n_{ct} = n_p/n_s$, where $n_s$ is the refractive index of the mixed solvent. A differential refractometer (C. N. Wood Mfg. Co., model RF-600) is used to measure $n_s$ at a wavelength of 546 nm and 20°C.

Because CHB is a better solvent for polystyrene than CH, increasing the percentage of CH in the solvent mixture causes a decrease in the radius of gyration, $R_g$, of the polystyrene depletant. The change in $R_g$ as a function of solvent quality consequently determines the depletion interaction due to its effect on $c^*$, the overlap concentration of the polystyrene which dictates the strength of the potential. Here, $c^* = 3M_w/(4\pi R_g^3 N_A)$, where $N_A$ is Avogadro's number. The values of $R_g$ and the second virial coefficient representing solvent quality, $A_2$, are measured using a multi-angle laser light scattering device (DAWN EOS, GaAs laser at 690 nm, Wyatt Technology) [Ganesan et al. (2013)]. The light scattered by the polystyrene at five different concentrations is analyzed with Zimm's theory [Podzimek (2011)]. Mixed solvent effects are negligible in the range of solvents tested [Solomon and Muller (1996)].

Table I. Physical properties of CH and CHB solvent mixtures.

| $w_{CH}$ | $w_{CHB}$ | $\rho_s$† (g/mL) | $\Delta\rho/\rho_p$ (%) | $Pe$ | $n_s$†† | $n_{ct}$††† | $R_g$ (nm) | $c/c^*$†††† |
|---|---|---|---|---|---|---|---|---|
| 0.10 | 0.90 | 1.243 | -4.6 | -0.060 | 1.501 | 1.007 | – | 1.33 |
| 0.16 | 0.84 | 1.190 | -0.2 | -0.002 | 1.493 | 1.012 | 37 ± 1 | 1.18 |
| 0.20 | 0.80 | 1.162 | 2.2 | 0.028 | 1.489 | 1.016 | 39 ± 2 | 1.09 |
| 0.28 | 0.72 | 1.105 | 7.0 | 0.091 | 1.480 | 1.021 | 33 ± 2 | 0.94 |
| 0.37 | 0.63 | 1.067 | 10.2 | 0.132 | 1.469 | 1.029 | 26 ± 3 | 0.81 |
| 0.47 | 0.53 | 1.049 | 11.7 | 0.152 | 1.461 | 1.035 | 31 ± 2 | 0.69 |
| 0.58 | 0.42 | 0.941 | 20.8 | 0.270 | 1.451 | 1.041 | 28 ± 2 | 0.58 |
| 0.64 | 0.36 | 0.912 | 23.2 | 0.301 | 1.447 | 1.045 | 32 ± 2 | 0.53 |
| 0.70 | 0.30 | 0.884 | 25.6 | 0.332 | 1.442 | 1.048 | 27 ± 2 | 0.48 |
| 0.84 | 0.16 | 0.829 | 30.2 | 0.392 | 1.434 | 1.054 | 25 ± 2 | 0.40 |

† measurements at 25°C
†† measurements at 546 nm, 20°C
††† shifted to 1046 nm, 20°C
†††† calculated from linear regression of $R_g$ as a function of $w_{CH}$

B. Optical trapping experiments

A 4W CW Nd:YVO$_4$ near-infrared (NIR, $\lambda$ = 1064 nm) laser (Compass 1064-4000M) is used as part of a custom-built laser tweezer set up described previously by Shindel et al. (2013). Briefly, two orthogonal acousto optic deflectors (AOD) are used to steer the laser position in the sample, and back-focal plane (BFP) interferometry is used to detect the displacement of the particles from the center of the trap based on the interference pattern of the scattered and unscattered light that pass out of the sample. By collimating the laser light after it passes through the sample and directing it onto a quadrant photodiode (QPD), we measure the lateral displacement of the particles from the traps in the $x$ and $y$ directions as shown by Gittes and Schmidt (1998).

To study the effect of index of refraction contrast on QPD sensitivity and trapping strength, dilute suspensions of PMMA particles in mixtures of CH and CHB are sealed in custom microscopy chambers made using a slide and coverslip spacers (No. 1.5) and sealed using a UV curable adhesive (Norland Optical Adhesive 81). A single particle is trapped and positioned ~35-40 μm into the sample and the trap is calibrated using a drag calibration technique [Meyer et al. (2006), Neuman and Block (2004)]. The microscope stage is moved at a constant velocity to impose a drag force on the particle. For a Newtonian solvent, the drag force is $F_{drag} = 6\pi\eta a v$, where $\eta$ is the solvent viscosity and $v$ is the solvent velocity around the particle. By varying $v$ and measuring the displacement with the QPD, we determine the relationship between applied force and particle displacement. In addition to the relationship between force and displacement, the QPD detector also has a response for large particle displacements. In order to measure the large displacements in the drag calibration, a QPD response curve is needed for each solvent composition. To generate the QPD response curves, the QPD signal is recorded for discrete displacements by moving the trap known distances using the AOD.

Thermal rupture forces are measured using two time-shared optical traps with a laser power of 215 mW measured at the back aperture to the objective. A pair of PMMA particles are trapped in a mixture of CH and CHB ($w_{CH}$ = 0.37) with $c$ = 8.64 mg/mL. One particle is held in a stationary trap while a second particle is brought into contact with the first by controlling the motion of the second trap. A triangle wave signal is used to control the motion of the second particle to ensure that the trap moves at a constant velocity at all points during the experiment. The displacement of the particle from the static trap upon the approach and retraction of the moving particle is averaged over 120 cycles and used to calculate the cumulative probability distribution of rupture forces by using trajectory averaging as described by Swan et al. (2012).

The particles are imaged with a CCD camera (Hitachi, KP-M1AN, 30 fps) using brightfield microscopy. The brightness intensity of the images is normalized to a reference image for comparison.

C. Microstructural characterization of gels

An inverted confocal microscope (Nikon A1Rsi) equipped with a resonant scanner head and a high-speed piezo stage is used to capture the 3D structure of the gels. Gels are loaded into custom-built glass capillaries with 300 μm spacers to match the gap used in the rheometer. The top and bottom of the capillaries are sealed using glass coverslips suitable for microscopy (0.17 mm thickness). The gels are set quiescently for 30 minutes before imaging. 3D image stacks are taken from the bottom coverslip (image dimensions: 42 μm × 42 μm × 30 μm, voxel dimensions: 83 nm × 83 nm × 83 nm) at a speed of 15 slices per second. After the initial image volume is captured, the sample is quickly flipped upside down and another stack of the same dimension is obtained from the top coverslip to assess the role of sedimentation on gel structure. Image processing is performed using an algorithm that uses a Gaussian mask to filter out digital noise and that identifies particle centroids based on their intensity maxima [Crocker and Grier (1996)]. The static error in the particle location is ± 20 nm in the $x$-$y$ plane and ± 31 nm in the $x$-$z$ plane as determined by Dibble et al. (2006).

D. Rheological characterization of gels

Oscillatory rheometry is performed on a stress-controlled rheometer (AR-G2, TA Instruments) at T = 25°C. Colloidal gels are loaded onto a Peltier plate and a stainless steel parallel plate geometry ($d$ = 6 cm, gap $h$ = 300 μm) is lowered to the gap distance while rotating at $\omega$ = 1 rad/s to minimize the formation of bubbles at the sample interface. A parallel plate geometry is chosen to minimize confinement effects and a correction factor for the nonhomogeneous strain rate is implemented [Soskey and Winter (1984)]. We analyze the strain-dependent linear elastic modulus for poly(ethylene oxide) standards ($Mw = 2\times10^6$ g/mol, 4 wt%) at $h$ = 50, 100, 150, 200, 300, 400, and 500 μm, and for gel samples at $h$ = 300 and 500 μm. The results show the absence of gap effects at $h > 150$ μm for this geometry. To reduce evaporation of the volatile solvent, a solvent trap is used. The absence of slip is verified by comparing results of a smooth fixture and a fixture with a sand-blasted surface (parallel plate geometry $d$ = 6 cm) [Buscall (2010)]. After sample loading, gels are pre-sheared unidirectionally for 1 minute (strain = 60,000, shear rate = 1000 s$^{-1}$). Oscillatory strain sweep (fixed angular frequency, $\omega$ = 10 rad/s) and frequency sweep measurements (fixed strain amplitude, $\gamma$ = 0.01) are performed after a waiting time of 30 minutes.

## III. RESULTS

A. Study Design and Solvent Composition Parameter Space

A successful model material should provide a compromise among the needs for refractive index contrast (for optical trapping), for refractive index matching (for microstructural characterization by confocal microscopy), and for density matching (for mechanical rheometry). The experimental requirements place major constraints on the model system: the refractive index contrast $n_{ct}$ must be greater than 1 for trapping, but should also be kept as low as possible to maintain resolution during 3D confocal imaging and to reduce the $\Delta\rho$ which introduces sedimentation effects into the rheology data.

Because the $n_{ct}$ required by optical trapping greatly increases the scattering of PMMA colloids deep within the specimen, images captured with confocal microscopy are limited to a depth of field of 30 μm in the $w_{CH}$ range in which trapping is possible. This depth of field is achieved up to the largest $n_{ct}$ at which confocal microscopy is performed ($w_{CH}$ = 0.64, $n_{ct}$ = 1.045). At $n_{ct} \sim 1$ ($w_{CH}$ = 0.10), the depth of field is limited only by the working distance of the objective (110 μm). The $n_{ct}$ requirement also means that a perfectly density matched system that is optimal for rheometry measurements ($w_{CH}$ = 0.16, $n_{ct} \sim 1$, $\Delta\rho/\rho_p \sim -0.002$) will not be compatible with optical trapping. To address these challenges and balance the conflicting constraints imposed on solvent composition for the three sets of measurements, the solvent composition is systematically varied, as per the systems reported in Table I, to search for a balance of physical properties in which bulk rheology, confocal microscopy, and laser tweezer measurements can all be performed on the same specimen.

B. Athermal limit of the interparticle potential

The derivative of the interparticle potential, $U(r)$, determines the maximum athermal rupture force, $f_{max}$. The analytical form of U(r) is obtained from the superposition of the Derjaguin-Landau-Verwey-Overbeek (DLVO) [Russel et al. (1989)] and Asakura-Oosawa (AO) theories [Asakura and Oosawa (1958)] for our model system. The electrostatic contributions are calculated using a Debye length of 1.3 μm (solvent conductivity for $w_{CH}$ = 0.37 mixture is $1.06 \times 10^{-8}$ S/m) and a constraint on the zeta potential of the solvent mixtures to be $\leq 10$ mV [Hsiao et al. (2012)]. The contact potential takes into consideration that the PHSA comb steric stabilizer is ~10 nm [Campbell and Bartlett (2002)]. We assume the stabilizer does not penetrate and use the slope of $U(r)$ evaluated at $r = 2a + 20$ nm to calculate $f_{max}$. At a constant polymer concentration ($c$ = 8.64 mg/mL), $f_{max}$ decreases with CH content because Rg and consequently $c/c^*$ decrease with a drop in solvent quality. Static light scattering measurements show that the second virial coefficient (a measure of polymer-solvent quality), $A_2$, decreases from $2.4 \times 10^{-6}$ to $1.5 \times 10^{-6}$ mL.mol/g$^2$ as CH concentration is increased from $w_{CH}$ = 0.16 to 0.84 [Fig. 1(a)], while the measured $R_g$ decreases from 37 nm to 25 nm [Fig. 1(b)]. Least-squares regression is used to fit the change in Rg (in nm) as a function of wCH with the form of $R_g$ = (-16 ± 4)$w_{CH}$ + (38 ± 2). This fit is used to compute the contribution of the depletion potential to $f_{max}$, the athermal force limit, which will be compared to the thermal rupture forces obtained from optical trapping.

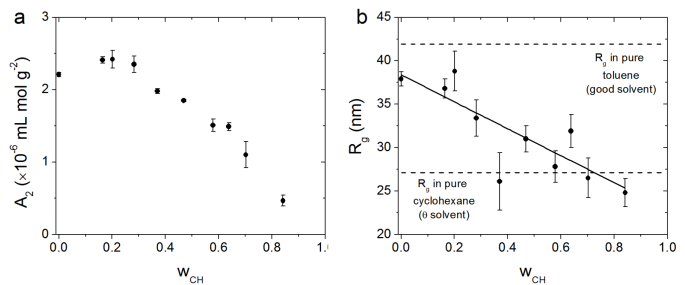

FIG. 1. Static light scattering of polystyrene in CH/CHB mixtures. (a) Changes in solvent quality, and (b) $R_g$ of 900,000 g/mol polystyrene in the solvents used in this study. In (b), the solid line is a least-squares fit of the change in $R_g$ as a function of $w_{CH}$. Literature values of $R_g$ in a good solvent (pure toluene) and a theta-solvent (pure CH) are provided for reference [Fetters, et al. (1994)].

C. Two-particle force measurements with laser tweezers

Direct measurements of the interactions between particles using optical tweezers rely on having sufficient refractive index contrast with the solvent to trap and simultaneously detect the particle positions. The reduction in refractive index contrast, $n_{ct}$, as the concentration of cyclohexane ($w_{CH}$) decreases can be qualitatively seen using bright field microscopy (Fig. 2). In this section, we characterize how the change in $n_{ct}$ affects both the trapping strength and the QPD position detector sensitivity.

First, we consider the position detector sensitivity. Gittes and Schmidt (1998) showed that the absolute detector response along a single axis depends on the refractive index contrast according to

$$\frac{I_+ - I_-}{I_+ + I_-} \sim \frac{n_s(n_{ct}^2 - 1)}{n_{ct}^2 + 2}, \quad (1)$$

where $I_+$ and $I_-$ are the scattering intensities in the positive and negative directions. As $n_{ct} \to 1$, the trapped particle scatters less light and the detector response decreases.

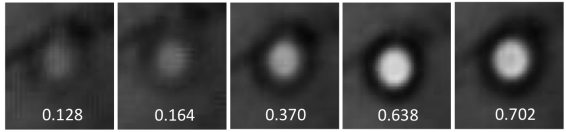

FIG. 2. Bright field microscopy images of PMMA particles ($2a$ = 2.7 μm) in mixtures of CH and CHB. Labels indicate the mass fraction of CH.

The QPD response curves for different solvent compositions are shown in Fig. 3(a). Data at each composition are fit with a polynomial to obtain a calibration curve for the tweezer experiments that converts the QPD signal to displacement. A Taylor series expansion around zero displacement describes the linear region of the QPD response. Comparing the magnitudes of the slopes for the different solvent compositions in Fig. 3(b) shows the QPD dependence on refractive index contrast. The inset on Fig. 3(b) shows the relationship between $n_{ct}$ and $w_{CH}$ as measured at 546 nm. The data is then shifted to represent $n_{ct}$ at the optical trapping condition (1064 nm) so that $n_{ct}$ = 1 at $w_{CH}$ = 3.5 ± 0.5% where we experimentally find that trapping becomes impossible. A higher contrast at a higher wavelength is reasonable because the permittivity of hydrocarbons increases with increasing wavelength in the near-IR band [Israelachvili (2011)]. The refractive index is approximately equal to the square root of the permittivity. As $w_{CH}$ is reduced, $n_{ct} \to 1$ and the sensitivity decreases, it is still possible to trap the particles, but the trapping conditions are marginal at $w_{CH}$ < 0.16 and the detector is not sensitive to small forces acting on the trapped particles.

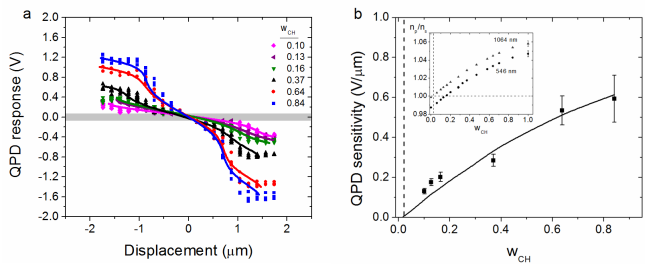

Figure 3. (a) QPD response to the displacement of PMMA particles ($2a$ = 2.7 μm) from an optical trap in mixtures of CH and CHB. The QPD response must be outside the gray box to be detectable. (b) QPD sensitivity as a function of $w_{CH}$. Inset is refractive index contrast as a function of $w_{CH}$ measured at 546 nm and shifted to represent contrast at trapping wavelength (1064 nm). The vertical dashed line marks the minimum $w_{CH}$ required for trapping.

Optical tweezers use a high numerical aperture objective to focus a laser to a diffraction-limited spot within the sample. The strength of the optical trap, and thus the magnitude of the forces that we can measure with the laser tweezers, is a function of $n_{ct}$. Trapping requires that $n_{ct}$ > 1. For small displacements from the trap, the force exerted on the particle is directly proportional to the displacement. The proportionality constant is the trap stiffness and is analogous to the spring constant of a Hookean spring. As the force increases, the relationship between force and displacement becomes nonlinear and the trapping strength cannot be represented as a constant, as shown in Fig. 4.

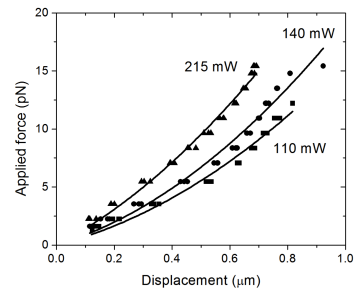

FIG. 4. Applied force as a function of displacement from drag calibrations at $w_{CH}$ = 0.84 as a function of laser power. There are three replicates at each laser power. Lines are quadratic fits to the data.

The drag calibration is performed for different solvent compositions ranging between $w_{CH}$ = 0.10 to 0.84. For reference, results for the drag calibration at $w_{CH}$ = 0.84 and laser powers of 110, 140, and 215 mW are shown in Fig. 4 (results from other $w_{CH}$ values are not shown). The first two terms of a Taylor series expansion around zero are used to calculate the trap stiffness, $k_T^0$, for the linear region. The single trap stiffness $k_T^0$ is validated using an alternative calibration method described by Shindel et al. (2013) that is called sequential impulse response (SIR). The SIR technique is valid for small displacements and does not require a priori knowledge about the size of the particle or the viscosity of the medium. The trapping strength increases with laser power and $n_{ct}$ (Fig. 5). Due to the low QPD sensitivity at $w_{CH}$ < 0.16, there are fewer data points composing the calibration curves and the error associated with is higher. Therefore, although the particles are nearly density matched for $w_{CH}$ < 0.16, they are poor conditions for accurately measuring interparticle forces.

In order to avoid making measurements based on large particle displacements where both the QPD response and the trapping force become highly nonlinear, we limit measurement displacements to within 2/3 of the particle radius. Evaluating the drag calibration curves at a displacement of $2a/3$ gives a conservative estimate for the maximum force that can be applied. This limit determines the operating regime of the force measurement experiment.

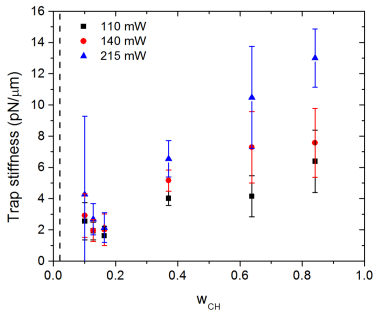

FIG. 5. Trap stiffness of PMMA particles ($2a = 2.7$ μm) in mixtures of CH and CHB as a function of $w_{CH}$ measured using the drag calibration method at 110, 140, and 215 mW. Error bars are 95% confidence intervals.

Using the measured trap stiffness from the drag calibration, we divide it in half to account for time-sharing the traps during the rupture force measurements. The resulting operating regime is shown in Fig. 6. To measure the thermal rupture forces with the laser tweezers, the maximum trapping force must be greater than $f_{max}$. This limit ensures that all rupture forces will be sampled. Therefore, low CH compositions impose a limit on the range of depletant polymer concentrations that can be studied. Fig. 6 suggests that with the maximum laser operating power of 215 mW, trapping measurements in the CH/CHB system should be possible at $w_{CH} \geq 0.16$ with the depletant range set by the CH composition. For example, at $w_{CH} = 0.16$, forces could be measured in systems up to $c \sim 3$ mg/mL, whereas at $w_{CH} = 0.47$, the depletant concentration can be nearly 16 mg/mL. For the particular case of $c = 8.64$ mg/mL ($c/c^* = 0.81$) studied in the present case, we find that at maximum trapping power, measurements can be performed when $w_{CH} > 0.30$.

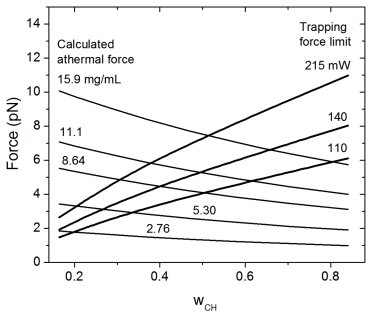

FIG. 6. Operating regime for rupture force measurements. The athermal force limit is calculated from the DLVO and Asakura-Oosawa potential. The maximum trapping force was determined from the drag calibration method for laser powers of 110, 140, and 215 mW.

D. Microstructural characterization

The cluster-like structures of gels produced in the CH/CHB solvent system are similar to those that have been reported for other systems with weak aggregation due to short-ranged attraction [Dibble et al. (2006)]. Figs. 7(a)-(c) show the representative images of quiescent gels in the $x$-$y$ plane, parallel to a bounding surface, where Fig. 7(a) corresponds to $w_{CH} = 0.16$ ($Pe = -0.002$), (b) corresponds to $w_{CH} = 0.37$ ($Pe = 0.132$), and (c) corresponds to $w_{CH} = 0.64$ ($Pe = 0.301$). These images are captured at $z = 10$ μm above the bottom of the sample to ameliorate potential wall effects on gel microstructure.

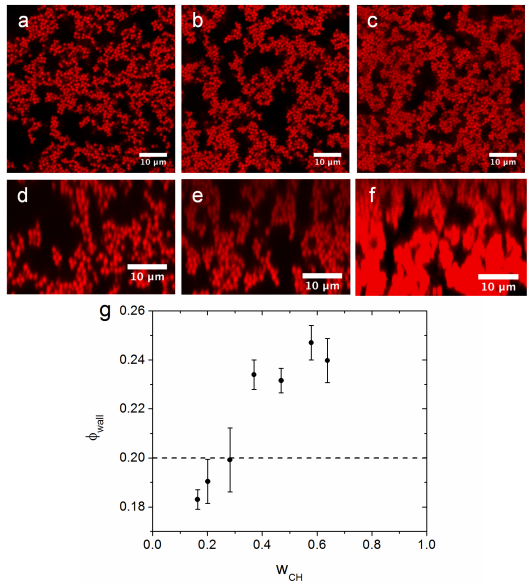

FIG. 7. Sedimentation of gels ($\phi_{gel} = 0.20$) at different solvent compositions. (a-c), Representative 2D $x$-$y$ plane CLSM images of gels with $w_{CH} = 0.16, 0.37$, and $0.64$; images are taken at $z = 10$ μm; (d-f), representative 2D $x$-$z$ plane ($z = 0$-$30$ μm) CLSM images of gels at the same solvent compositions; (g), $\phi_{wall}$ as a function of $w_{CH}$ estimated from the number of particles in the $x$-$y$ plane at $z = 10$ μm. Error bars shown are standard deviations of the mean.

When $w_{CH} \geq 0.37$, the effects of sedimentation on gel structure are readily observed from $x$-$z$ images. Fig. 7(d)-(f) show representative images of the gels in the $x$-$z$ plane (the $z$ dimension origin is defined as the bottom; $z = 300$ μm is the top coverslip). For example, compaction of the bottom layer from sedimentation is significant at $w_{CH} = 0.64$ [Fig. 7(f)]. This compaction is quantified by measurement of the near-wall volume fraction, $\phi_{wall}$, which is defined as the number of particles found in a thin region centered at $z = 10$ μm, divided by the volume of that region given by the length and width of the image and a thickness equal to the particle diameter (42 μm × 42 μm × 1.15 μm). For comparison, the homogeneous (as prepared) volume fraction of these specimens is $\phi_{gel} = 0.20$, and this value would be expected for the sedimentation-free condition [Guo and Lewis (2000)]. Fig. 7(g) shows the change in $\phi_{wall}$ as a function of the amount of CH in the mixture, where an increasing amount of CH results in a larger $\Delta\rho/\rho$ and $Pe$. At $w_{CH}$ = 0.37 ($Pe$ = 0.132), the value of $\phi_{wall}$ increases to 0.234 ± 0.006 and supports the visual observations of compaction through sedimentation. Experimental error of $\phi_{wall}$ from instrument limitations and image processing are ~ 10% of the mean value. Thus, we consider values of $\phi_{wall} \geq 0.22$ to be cases where a significant amount of compaction has arisen due to sedimentation.

This type of compaction is reminiscent of delayed sedimentation in weakly aggregated gels with short-ranged attractions [Sedgwick et al. (2004)] but in our case we observe a sudden, rapid settling at a fixed waiting time at $w_{CH} \geq 0.37$ as opposed to a critical sedimentation time scale. It is interesting to note that the Péclet number in our case is quite high, but a stable gel structure is maintained at a fixed waiting time of 30 minutes with minimal sedimentation effects on the microstructure. For example, recent literature for thermoreversible adhesive hard spheres report a critical gravitational Péclet number, $Pe_g$, of 0.01 or less for stable gels [Kim et al. (2013)]. This Péclet number also compares sedimentation to Brownian diffusion with added effect of percolation in gel networks. (We do not work with fractal gels in our system.) Regardless, when we substitute system-specific variables and assume a DLCA limit for fractal gels ($d_f \sim 1.8$) [Allain et al. (1995)], we find that the $Pe_g$ of the system at $w_{CH}$ = 0.37 is ~ 4, which is far greater than the critical criterion for stability. This suggests that delayed sedimentation does occur in our system, but not at the time scale of the experiment for samples at $w_{CH} \leq 0.37$.

E. Linear and nonlinear rheology of gels

Colloidal gels produced in solvents of different wCH display variable rheology that depends on a number of factors: (i) sedimentation propensity due to differences in the degree of density matching; (ii) van der Waals attraction due to differences in the dielectric spectra of CH and CHB; (iii) depletion attraction due to solvent-quality induced change in the radius of gyration of polystyrene. Specifically, as a result of the decrease in $n_{ct}$ when $w_{CH}$ decreases, the van der Waals attraction between particles decreases. However, the van der Waals interaction at contact is negligible (< $10^{-4}$ pN at all rheological experimental conditions) compared to the change in the depletion attraction caused by the increase in $R_g$ (~100 pN). Gravitational sedimentation is an additional factor that complicates the understanding of the rheological phenomenon. Here, we use information gained from confocal microscopy to delineate the region in which sedimentation is the dominant factor contributing to the rheological phenomena.

The linear viscoelasticity of the gels is reported as a function of frequency [Fig. 8] for different $w_{CH}$. The slopes of the frequency-dependent elastic and viscous moduli, $G'(\omega)$ and $G''(\omega)$, are characteristic of the solid-like properties of the gels [Larson (1999)]. For viscoelastic samples, a large slope corresponds to liquid-like behavior.

At the gel point, the samples become sufficiently solid-like such that $G'(\omega)$ and $G''(\omega)$ should only be weakly dependent on ω [Chambon and Winter (1987); Rueb and Zukoski (1998)]. The frequency sweep data shows that gels become weaker in strength as $w_{CH}$ increases, consistent with the decrease in both $c/c^*$ and $R_g$, which affect the strength of the depletion attraction (see Table I).

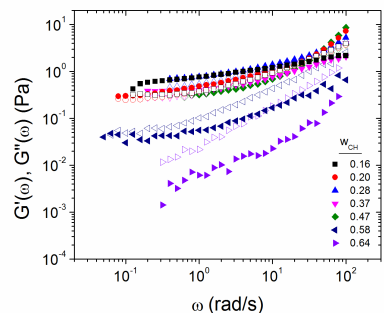

FIG. 8. Linear oscillatory frequency sweep dataset for gels with $w_{CH}$ = 0.16, 0.20, 0.28, 0.37, 0.47, 0.58, and 0.64. Closed symbols represent G' and open symbols represent G".

In Fig. 8, sedimentation effects do contribute to the decrease in the frequency-dependent moduli, but these effects cannot be easily distinguished from those due to the solvent-quality induced change in attraction strength. The strain-dependent viscoelasticity of the gels provides a much clearer picture [Fig. 9] of the role of sedimentation. A gradual decrease in the plateau value, G', is observed from $w_{CH}$ = 0.16 to 0.37 [Fig. 10(a), from (1) to (4)]. For $w_{CH}$ > 0.37 [Fig. 10(a), from (4) to (5)], G' decreases much more rapidly. This sharp decline in the elastic modulus can be attributed to gel detachment from the rheometer geometry, consistent with observations of compaction [Fig. 10(b-k)]. This correspondence suggests that at conditions of large density mismatch ($Pe$ > 0.132), the gel strength is insufficient to maintain sample integrity, and the collapse of the gel results in a fluid layer at the top of the sample and compaction of particles at the bottom [Figs. 10(j) and (k)].

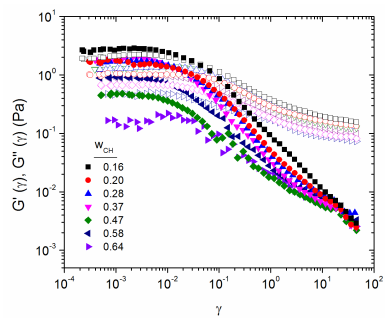

FIG. 9. Linear oscillatory strain sweep dataset for gels with $w_{CH}$ = 0.16, 0.20, 0.28, 0.37, 0.47, 0.58, and 0.64. Closed symbols represent G' and open symbols represent G".

This fluid region acts as a slip layer in the rheometer that reduces the elastic modulus in a manner that is far more significant than the decrease in gel strength due to the decrease in depletion attraction [Mewis and Wagner (2012)]. We therefore conclude that of the three potential contributions to the variability in G' with $w_{CH}$, the effect of density mismatch is the most significant when $w_{CH} > 0.37$. The sedimentation effects limit measurements of rheology in this system to $w_{CH} \leq 0.37$ for the particular case of depletion concentration $c = 8.64$ mg/mL studied here.

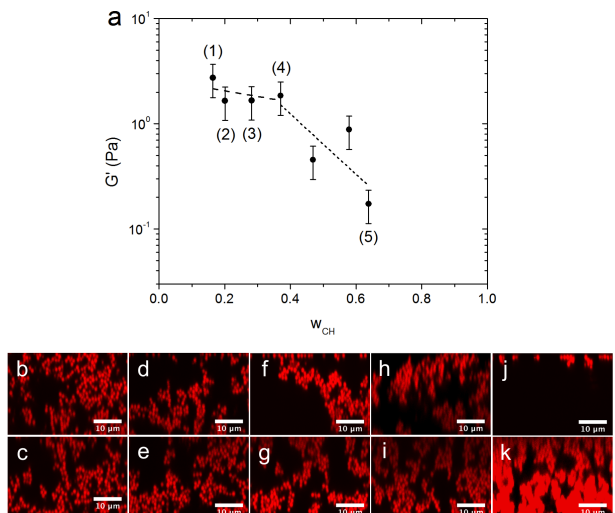

FIG. 10. Visualization of sedimentation and effect on linear rheology. (a) Linear strain-dependent G' as a function of $w_{CH}$. Dashed and dotted lines separate the two different regimes. Error bars shown are standard errors of the mean from seven independent measurements of one sample at $w_{CH} = 0.64$. Representative 2D x-z confocal images of gels at (b,c), $w_{CH} = 0.16$, (d,e), $w_{CH} = 0.20$, (f,g), $w_{CH} = 0.28$, (h,i), $w_{CH} = 0.37$, and (j,k) $w_{CH} = 0.64$. (b-j) show the z = 270 to 300 μm and (c-k) show z = 0 to 30 μm for the gels. In (a), (1) refers to the measured rheology at a gelation condition corresponding to (b,c), (2) = (d,e), (3) = (f,g), (4) = (h,i), and (5) = (j,k).

## IV. DISCUSSION

Although maximizing the refractive index contrast $n_{ct}$ enhances trapping strength and therefore allows characterization of thermally induced rupture, the resultant refractive index and density mismatch complicates confocal microscopy and rheology experiments. The solvent that is density matched with PMMA ($w_{CH} = 0.16$), and therefore optimal for confocal microscopy and rheology, has very low index contrast ($n_{ct} = 1.007$) that reduces the resolution of the force measurements with the laser tweezers to unacceptable levels. We show that the detector sensitivity is significantly reduced at $w_{CH} \leq 0.16$ as $n_{ct} \to 1$ with a decrease in trapping strength. The maximum trapping strength sets the range of $w_{CH}$ where forces can be measured for a selected depletant concentration. For $c = 8.64$ mg/mL, interparticle forces can be measured at $w_{CH} > 0.30$, whereas compaction from sedimentation occurs at $w_{CH} \geq 0.37$, complicating microscopy and rheological measurements. The intersection of these two constraints indicates that $w_{CH} = 0.37$ is an optimal condition at which there is sufficient optical trapping strength and detector sensitivity for two-particle force measurements with minimal deleterious effects on microstructural characterization and rheological measurement.

A complete characterization of rupture force, gel microstructure, and linear viscoelasticity using this new gel system involving CH and CHB is therefore possible at $w_{CH} = 0.37$ and $c = 8.64$ mg/mL (Fig. 11). This particular gel ($\phi_{gel} = 0.20$, $c/c^* = 0.81$) has a cluster-like structure that is typical of weak depletion gels [Fig. 11(a)] [Dibble et al. (2006)], as seen in a broad contact number distribution, $p(z)$, with a mean contact number, $\langle z \rangle$, of $5.39 \pm 0.06$ [Fig. 11(b)]. The solvent of the gel has a refractive index contrast of $n_{ct} = 1.029$, which is sufficiently high for measuring interparticle forces. The cumulative probability distribution of a pair of particles being bonded as a function of applied force is shown in Fig. 11(c). At low applied forces, there is a high probability that the particles will still be bonded, but the distribution decays for higher applied forces. For a normal distribution, the force at which 50% of the bonds have broken, $f_{50}$, corresponds to the peak in the probability density function and is the most frequently measured rupture force. The $f_{50}$ value calculated from the rupture force distribution for this gel is $3.3 \pm 0.5$ pN. The forces where 5% and 95% of the bonds are still bonded set a range of forces describing the width of the distribution. The high end of that distribution is $f_{05} = 4.8 \pm 0.9$ pN, which is approximately equal to the calculated athermal rupture force ($f_{max} = 4.6 \pm 0.4$ pN), demonstrating that we have the ability to measure the maximum possible pair bonding force expected in this gel. While the particle sizes need to be smaller for resolvable rheological characterization, the difference between the viscoelasticity and measured thermal forces can be reconciled through the effects of particle diameter on depletion and Brownian motion [Mewis and Wagner (2012)].

The linear and nonlinear rheological properties of the gel are shown in Figs. 11(d)-(f). Fig. 11(d) shows the frequency-dependent elastic and viscous moduli, G'($\omega$) and G"($\omega$). The moduli are weakly dependent on $\omega$, which is typical for gels with arrested dynamics. Fig. 11(d) shows that G'($\gamma$) and G"($\gamma$) cross over at the rheological yield strain, $\gamma_c = 0.016$. The yield strain indicates a critical point after which further deformation results in a significant fluidization of the sample. The yield stress, defined by $\tau_y = \gamma G'$ [Yang et al. (1986); Koumakis and Petekidis (2011)], shows a first peak at the onset of nonlinear elastic response [Fig. 11(f)], which is typically seen for colloidal gels undergoing bond breaking [Koumakis and Petekidis (2011)]. For this particular gel, the value of $\tau_y$ is $0.04 \pm 0.01$ Pa. We do not observe a distinct second yield strain that corresponds to cage breaking in this sample [Koumakis and Petekidis (2011)].

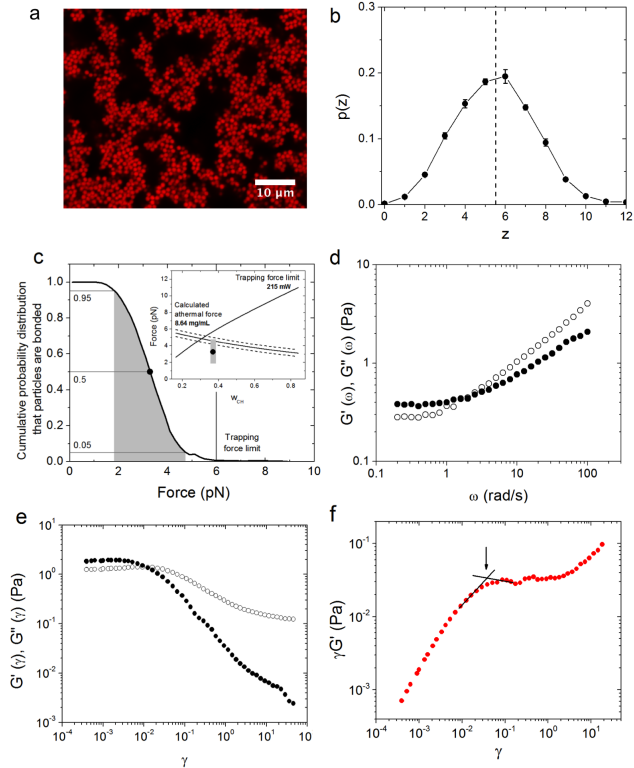

FIG. 11. Characterization of a model colloidal gel ($w_{CH}$ = 0.37, $c$ = 8.64 mg/mL, 2a = 1.15 μm). (a) 2D *x-y* image of the gel at *z* = 10 μm; (b) 3D contact number distribution $p(z)$, where the dotted line is the mean contact number, ⟨z⟩; (c) Cumulative probability distribution of rupture forces between PMMA particles (2a = 2.7 μm) calculated from 120 rupture cycles. The highlighted forces mark the range in which 5% and 95 % of bonds are bonded. The inset shows where the force distribution fits on the trapping operating regime. The dashed lines mark the 95% confidence intervals around the athermal limit due to uncertainty in $R_g$. (d) Frequency-dependent viscoelasticity and (e) strain-dependent viscoelasticity. Closed symbols represent G'(ω) and G'(γ) and open symbols represent G"(ω) and G"(γ). (f) Elastic yield stress of gel with the arrow indicating the first yield stress, $\tau_y$.

## V. CONCLUSIONS

We have introduced a new model depletion gel in which the interparticle forces, microstructure, and rheology can all be measured. The system uses CH and CHB as the solvent mixture so as to balance the constraints imposed by optical trapping, confocal microscopy, and rheometry. At a weight ratio of $w_{CH}$ = 0.37, the refractive index contrast ($n_{ct}$ = 1.029) is sufficient to allow measurement of forces as high as 6-7 pN, yet low enough to enable confocal imaging of the microstructure. The density difference between the PMMA and the solvent ($\Delta\rho/\rho_p$ = 0.102) is sufficiently small that sedimentation does not significantly impact rheological measurements at short experimental time scales.

This new system serves as a tool for studying the relationship between microscopic properties and the rheology of depletion gels. Current micromechanical models rely on the athermal force limit calculated from theory to predict yielding behavior. Our force measurements show that there is a distribution of rupture forces that result in bond breakage. The way to account for the force distribution in order to explain yielding behavior or microstructure rearrangements is not yet available; however, measurements of the kind described here are a precondition to generating such understanding. Although the bond strength is directly related to the yield stress in colloidal gels, the consequences of the coupling between the distribution of interparticle forces and the collective gel structure is less clear for other linear and nonlinear rheological properties. This model system will enable future exploration of the connection between the microscopic interparticle forces and linear and nonlinear rheology of colloidal gels with short-ranged attractions.

**ACKNOWLEDGEMENTS**
We are grateful to J. Swan and M. Shindel for help with the optical tweezers and useful discussions on the stochastic nature of rupture forces. This work is supported by the International Fine Particles Research Institute (IFPRI) and the National Science Foundation (NSF) under awards CBET-1235955 and CBET-0853648.

† These authors contributed equally to this work.
* Corresponding author: furst@udel.edu

**REFERENCES**
1. Allain, C., M. Cloitre and M. Wafra, "Aggregation and sedimentation in colloidal suspensions", Phys. Rev. Lett., **74**, 1478 (1995).
2. Antl, L., J. W. Goodwin, R. D. Hill, R. H. Ottewill, S. M. Owens, S. Papworth, and J. A. Waters, "The preparation of poly(methyl methacrylate) lattices in non-aqueous media," Colloids Surf, **17**, 67-78 (1986).
3. Asakura, S. and F. Oosawa, "Interaction between particles suspended in solutions of macromolecules," J. Polym. Sci., **33**, 183 (1958).
4. Buscall, R., "Letter to the editor: Wall slip in dispersion rheometry", J. Rheol., **54**, 1177 (2010).
5. Campbell, A. I., and P. Bartlett, "Fluorescent hard-sphere polymer colloids for confocal microscopy," J. Coll. Int. Sci., **256**, 325-330 (2002).

6. Chambon, F. and H. H. Winter, "Linear viscoelasticity at the gel point of a crosslinking PDMS with imbalanced stoichiometry," J. Rheol., **31**, 683 (1987).
7. Chan, H. and A. Mohraz, "Two-step yielding and directional strain-induced strengthening in dilute colloidal gels", Phys. Rev. E, 85, 041403 (2012).
8. Crocker, J. C. and D. G. Grier, "Methods of digital video microscopy for colloidal studies," J. Coll. Inter. Sci., **179**, 298-310 (1996).
9. Crocker, J. C., J. A. Matteo, A. D. Dinsmore, and A. G. Yodh, "Entropic attraction and repulsion in binary colloids probed with a line optical tweezer, Phys. Rev. Lett., **82**, 4352-4355 (1998).
10. Dibble, C. J., M. Kogan, and M. J. Solomon, "Structure and dynamics of colloidal depletion gels: Coincidence of transitions and heterogeneity," Phys. Rev. E, **74**, 041403 (2006).
11. Drury, J. L. and D. J. Mooney, "Hydrogels for tissue engineering: scaffold design variables and applications," Biomaterials, **24**, 4337-4351 (2003).
12. Emady, H, M. Caggioni, and P. Spicer, "Colloidal microstructure effects on particle sedimentation in yield stress fluids," J. Rheol., **57**, 1761 (2013).
13. Evans, E., "Introductory lecture energy landscapes of biomolecular adhesions and receptor anchorin at interfaces explored with dynamic force spectroscopy," Faraday Discuss., **111**, 1-16 (1998).
14. Fetters, L. J., N. Hadjichristidis, J. S. Lindner and J. W. Mays, "Molecular weight dependence of hydrodynamic and thermodynamic properties for well-defined linear polymers in solution," J. Phys. Chem. Ref. Data, **23**, 619-640 (1994).
15. Fielding, S. M., P. Sollich and M. E. Cates, "Aging and rheology in soft materials", J. Rheol., **44**, 323 (2000).
16. Furst, E. M. and J. P. Pantina, "Yielding in colloidal gels due to nonlinear microstructure bending mechanics," Phys. Rev. E, **75**, 050402(R), (2007).
17. Ganesan, M., E. J. Stewart, J. Szafranski, A. E. Satorius, J. G. Younger, and M. J. Solomon, "Molar mass, entanglement, and associations of the biofilm polysaccharide of Staphylococcus epidermidis," Biomacromolecules, **14**, 1474 (2013).
18. Gittes, F. and C. F. Schmidt, "Interference model for back-focal-plane displacement detection in optical tweezers," Opt. Lett., **23**, 7-9 (1998).
19. Gopalakrishnan, V. and C. F. Zukoski, "Delayed flow in thermo-reversible colloidal gels", J. Rheol., **51**, 623 (2007).
20. Guo, J. and J. A. Lewis, "Effects of ammonium chloride on the rheological properties and sedimentation behavior of aqueous silica suspensions", J. Am. Ceram. Soc., **83**, 266 (2000).
21. Helgeson, M. E., S. E. Moran, H. Z. An, and P. S. Doyle, "Mesoporous organohydrogels from thermogelling photocrosslinkable nanoemulsions," Nature Materials, **11**, 344-352 (2012).
22. Hess, A. and N. Aksel, "Yield stress and scaling of polyelectrolyte multilayer modified suspensions: Effect of polyelectrolyte conformation during multilayer assembly", Langmuir, **29**, 11236 (2013).
23. Hoekstra, H., J. Vermant, J. Mewis, G. G. Fuller, "Flow-induced anisotropic and reversible aggregation of two-dimensional suspensions," Langmuir, **19**, 9134-9141 (2003).
24. Hsiao, L. C., R. S. Newman, S. C. Glotzer and M. J. Solomon, "Role of isostaticity and load-bearing microstructure in the elasticity of yielded colloidal gels," PNAS, **109**, 16029-16034 (2012).
25. Huh, J. Y., M. L. Lynch, and E. M. Furst, "Microscopic structure and collapse of depletion-induced gels in vesicle-polymer mixtures," Phys. Rev. E **76**, 051409 (2007).
26. Israelachvili, J. N., "Intermolecular and surface forces", Academic Press, 3$^{rd}$ edition, 2011.
27. Kim, J. M., J. Fang, A. P. R. Eberle, R. Castañeda-Priego and N. J. Wagner, "Gel transition in adhesive hard-sphere colloidal dispersions: The role of gravitational effects", Phys. Rev. Lett., **110**, 208302 (2013).
28. Koumakis, N. and G. Petekidis, "Two step yielding in attractive colloids: transition from gels to attractive glasses," Soft Matter, **7**, 2456-2470 (2011).
29. Kramers, H. A., "Brownian motion in a field of force and the diffusion model of chemical reactions," Physica, **7**, 284-304 (1940).
30. Larson, R. G., "The structure and rheology of complex fluids," (1999).
31. Lee, M. H. and E. M. Furst, "Response of a colloidal gel to a microscopic oscillatory strain," Phys. Rev. E, **77**, 041408 (2008).
32. Lu, Y., R. Ganguli, C. A. Drewien, M. T. Anderson, C. J. Brinker, W. Gong, Y. Guo, H. Soyez, B. Dunn, M. H. Huang, and J. I. Zink, "Continuous formation of supported cubic and hexagonal mesoporous films by sol-gel dip-coating," Nature, **389**, 364-368 (1997).
33. Masschaele, K., J. Fransaer and J. Vermant, "Flow-induced structure in colloidal gels: direct visualization of model 2D suspensions", Soft Matter, **7**, 7717 (2011).
34. Manley, S., L. Cipelletti, V. Trappe, A. E. Bailey, R. J. Christianson, U. Gasser, V. Prasad, P. N. Segre, M. P. Doherty, S. Sankaran, A. L. Jankovsky, B. Shiley, J. Bowen, J. Eggers, C. Kurta, T. Lorik and D. A. Weitz, "Limits to gelation in colloidal aggregation", Phys. Rev. Lett., **93**, 108302 (2004).
35. Maranzano, B. J. and N. J. Wagner, "The effects of interparticle interactions and particle size on reversible shear thickening: Hard-sphere colloidal dispersions", J. Rheol., **45**, 1205 (2001).
36. Mewis, J. and N. J. Wagner, "Colloidal suspension rheology," (2012).
37. Neuman, K. C. and S. M. Block, "Optical trapping," Rev. Sci. Instrum., **75**, 2787-2809 (2004).

38. Pantina, J. P. and E. M. Furst, "Elasticity and critical bending moment of model colloidal aggregates," Phys. Rev. Lett., **94**, 138301 (2005).
39. Pantina, J. P. and E. M. Furst, "Colloidal aggregate micromechanics in the presence of divalent ions," Langmuir, **22**, 5282-5288 (2006).
40. Pantina, J. P. and E. M. Furst, "Micromechanics and contact forces of colloidal aggregates in the presence of surfactants," Langmuir, **24**, 1141-1146 (2008).
41. Park, J. D. and K. H. Ahn, "Sturctural evolution of colloidal gels at intermediate volume fraction under start-up of shear flow", Soft Matter, **9**, 11650 (2013).
42. Podzimek, S., "Light scattering, size exclusion chromatography and asymmetric flow field flow fractionation" John Wiley & Sons, Inc.: New York, 2011; pp 207−258.
43. Potanin, A. A., "On the computer simulation of the deformation and breakup of colloidal aggregates in shear flow," J. Coll. Int. Sci., **157**, 399-410 (1998).
44. Rajaram, B. and A. Mohraz, "Microstructural response of dilute colloidal gels to nonlinear shear deformation", Soft Matter, **6**, 2246 (2010).
45. Ramakrishnan, S., V. Gopalakrishnan and C. F. Zukoski, "Clustering and mechanics in dense depletion and thermal gels", Langmuir, **21**, 9917 (2005).
46. Reddy, N. K., Z. Zhang, M. P. Lettinga, J. K. G. Dhont, and J. Vermant, "Probing structure in colloidal gels of thermoreversible rodlike virus particles: Rheology and scattering," J. Rheol., **56**, 1153 (2012).
47. Rich, J. P., J. Lammerding, G. H. McKinley and P. S. Doyle, "Nonlinear microrheology of an aging, yield stress fluid using magnetic tweezers", Soft Matter, **7**, 9933 (2011).
48. Royall, C. P., M. E. Leunissen and A. van Blaaderen, "A new colloidal system to study long-range interactions quantitatively in real-space", J. Phys.: Condens. Matter, 15, S3581 (2003).
49. Rueb, C. J. and C. F. Zukoski, "Rheology of suspensions of weakly attractive particles: Approach to gelation," J. Rheol., **42**, 1451 (1998).
50. Russel, W. A., D. A. Saville, and W. R. Schowalter, "Colloidal Dispersions" (1989).
51. Schall, P., D. A. Weitz, and F. Spaepen, "Structural rearrangements that govern flow in colloidal glasses," Science, **318**, 1895 (2007).
52. Sedgwick, H., S. U. Egelhaaf and W. C. K. Poon, "Clusters and gels in systems of sticky particles", J. Phys.: Condens. Matter, **16**, S4913 (2004).
53. Shindel, M. M., J. W. Swan, and E. M. Furst, "Calibration of an optical tweezer microrheometer by sequential impulse response," Rheol. Acta, **52**, 455-465 (2013).
54. Solomon, M. J. and S. J. Muller, "Study of mixed solvent quality in a polystyrene-dioctyl phthalate-polystyrene system", J. Poly. Sci.: Part B: Poly. Phys., **34**, 181 (1996).
55. Soskey, P. R. and H. H. Winter, "Large step shear strain experiments with parallel-disk rotational rheometers", J. Rheol., 28, 625 (1984).
56. Sriram, I., E. M. Furst, R. J. DePuit and T. M. Squires, "Small amplitude active oscillatory microrheology of a colloidal suspension", J. Rheol., **53**, 357 (2009).
57. Studart, A. R., E. Amstad and L. J. Gauckler. Yielding of weakly attractive nanoparticle networks. Soft Matter, 7(14), 6408-6412 (2011).
58. Swan, J. W., M. M. Shindel, and E. M. Furst, "Measuring thermal rupture force distributions from an ensemble of trajectories," Phys. Rev. Lett., **109**, 98302 (2012).
59. Varadan, P. and M. J. Solomon, "Direct visualization of flow-induced microstructure in dense colloidal gels by confocal laser scanning microscopy," J. Rheol., **47**, 943 (2003).
60. Vermant, J. and M. J. Solomon, "Flow-induced structure in colloidal suspensions", J. Phys.: Condens. Matter, **17**, R187 (2005).
61. Vermant, J., P. Van Puyvelde, P. Moldenaers and J. Mewis, "Anisotropy and orientation of the microstructure in viscous emulsions during shear flow", Langmuir, **14**, 1612 (1998).
62. West, A. H. L., J. R. Melrose and R. C. Ball, "Computer simulations of the breakup of colloidal aggregates", Phys. Rev. E, **49**, 4237 (1994).
63. Yang, M. C., L. E. Scriven and C. W. Macosko, "Some rheological measurements on magnetic iron oxide suspensions in silicone oil", J. Rheol., **30**, 1015 (1986).
64. Yin, G. and M. J. Solomon, "Soft glassy rheology model applied to stress relaxation of a thermoreversible colloidal gel", J. Rheol., **52**, 785 (2008).
65. Zaccarelli, E., "Colloidal gels: Equilibrium and non-equilibrium routes", J. Phys.: Condens. Matter, **19**, 323101 (2007).